\documentclass[twocolumn,trackchanges]{aastex7}

\newcommand{\mbh}{$M_{\rm BH}$}

\newcommand{\msun}{$M_{\odot}$}
\newcommand{\lsun}{$L_{\odot}$}

\usepackage{hyperref}
\usepackage{url}
\usepackage{xcolor}
\usepackage{gensymb}
\usepackage{mathtools}
\usepackage{siunitx}
\DeclarePairedDelimiterXPP\BigOSI[2]%
  {\mathcal{O}}{(}{)}{}%
  {\SI{#1}{#2}}

\let\oldmaketitle\maketitle
\renewcommand{\maketitle}{\oldmaketitle\setcounter{footnote}{0}}

\begin{document}

\title{Revisiting the Stellar-Dynamical Mass of the Central Black Hole in NGC 3227}

\author[0000-0002-6870-6144]{Nabanita Das}
\affiliation{Department of Physics and Astronomy,
		 Georgia State University,
		 Atlanta, GA 30303, USA}
\email{ndas5@gsu.edu}

\author[0000-0002-2816-5398]{Misty C.\ Bentz}
\affiliation{Department of Physics and Astronomy,
		 Georgia State University,
		 Atlanta, GA 30303, USA}
\email{bentz@astro.gsu.edu}

\author[0000-0002-5038-9267]{Eugene Vasiliev}
\affiliation{University of Surrey,
        Guildford, GU2 7XH, UK}
\email{eugvas@protonmail.com}

\author[0000-0002-6257-2341]{Monica Valluri}
\affiliation{Department of Astronomy,
         University of Michigan,
         Ann Arbor, MI, 48104, USA}
\email{mvalluri@umich.edu}

\author[0000-0003-0017-349X]{Christopher A.\ Onken}
\affiliation{Research School of Astronomy and Astrophysics, 
        Australian National University, 
        Canberra, ACT 2611, Australia}
\email{christopher.onken@anu.edu.au}

\defcitealias{davies06}{Davies et al.}
\defcitealias{bentz23b}{Bentz et al.}
\defcitealias{hicks08}{Hicks \& Malkan}

\begin{abstract}
We present a new stellar-dynamical mass for the supermassive black hole in the nearby Seyfert galaxy NGC 3227. We determined the line-of-sight velocity distributions of stars in the nuclear region of NGC 3227 using archival spatially-resolved $K$-band spectra obtained with NIFS at Gemini Observatory. The surface brightness profile of the galaxy is described by four components, measured from a WFC3 F547M image from the Hubble Space Telescope. We constructed dynamical models using the Schwarzschild orbit-superposition method implemented in the open-source code \texttt{FORSTAND}. We find a best-fit black hole mass of \mbh=$(1.2\pm0.5) \times 10^7$ \msun. This mass is consistent with a previous mass from \citeauthor{davies06} based on stellar dynamics using orbit-based modeling, but our results make use of higher signal-to-noise data from which the higher-order moments of the stellar LOSVDs are constrained, thus providing a more precise \mbh. Our result is also consistent with the \mbh\ constraint from velocity-resolved reverberation mapping reported by \citeauthor{bentz23b}.  And while our best-fit mass is slightly smaller than the mass constraint from gas-dynamical modeling reported by \citeauthor{hicks08} (after scaling to the same adopted distance), they are within 2$\sigma$ of each other.  Direct comparisons of black hole masses from different techniques are rare but crucial to ensure all black hole masses are on the same scale, and to identify and minimize the uncertainties and possible sources of bias associated with each technique.
\end{abstract}


\section{Introduction}\label{sec:intro}
Active galaxies are excellent targets for studying the co-evolution of supermassive black holes (SMBHs) and their host galaxies through the feedback of the active black hole on its host \citep{crenshaw12,kormendy13, falcone24}. One key component in understanding this connection is the mass of the central black hole (\mbh). However, there are only a handful of techniques available for us to directly measure the black hole mass.

One of the most widely used methods for SMBH mass measurement is dynamical modeling of the centers of their host galaxies, using stars or gas \citep{kormendy95, macchetto97, vandermarel98, walsh13, barth16, liepold23}. Both stellar-dynamical (SD) and gas-dynamical (GD) modeling require very high spatial resolution to map the bulk motions of the stars or gas as a function of position. Since spatial resolution degrades with distance, these methods are only applicable to galaxies in the local Universe (few tens Mpc).

In the case of active galactic nuclei (AGN), which are rare in the local universe, the most common mass measurement method is reverberation mapping (RM; \citealt{blandford82, peterson93, cackett21}). In this technique, the time delay between variations in the continuum emission and the response to those variations from the broad-line region (BLR) of an AGN is measured providing the light-travel time (or distance) between the two regions; by combining the time delay with the velocity of the broad-line gas, \mbh\ is constrained. This method does not require high spatial resolution
and can be applied to more distant galaxies. 

All \mbh\ measurement techniques  have underlying assumptions and may be subject to biases. Their selection criteria are also different, creating minimal overlap between SD and RM samples. Furthermore, AGN are often excluded from SD modeling studies  because the bright emission from the accreting black hole dilutes and may overwhelm the stellar signatures in the spectra.
As a result, we have two largely independent, differently calibrated SMBH mass scales --- one within the local neighborhood dominated by dynamical black hole masses, and one outside that is based on RM results.
In order to assess whether the masses measured using these two methods are consistent with each other, one must directly the compare masses derived from both techniques in the same galaxies. However, few broad-lined AGNs are available at distances where dynamical modeling methods may be used to constrain the black hole mass.

NGC 3227, as a nearby Type-1 Seyfert galaxy at redshift $z=0.00377$, is an ideal candidate for such a comparison study. It was included by Carl Seyfert in his original sample of 12  `extragalactic nebulae' \citep{seyfert43}.
Further study of NGC 3227 by \citet{rubin68} revealed several broad emission lines generated from fast moving gas clouds, indicating some short-timescale `explosive phenomena' at the nucleus of the galaxy. NGC 3227 is also known to be interacting with its companion NGC 3226, which is another active galaxy identified as a low-ionization nuclear emission-line region (LINER, \citealt{ho97,george01}).

The SMBH mass in NGC 3227 has been measured using different methods, including reverberation mapping \citep{denney10, derosa18, brotherton20, bentz23b}, stellar-dynamical modeling \citep{davies06}, and gas-dynamical modeling \citep{hicks08}.
To date, there are only three other active galaxies that have direct masses from RM and other \mbh\ measurement techniques, allowing detailed \mbh\ comparisons: NGC 4151 \citep{bentz06b, roberts21, bentz22}, NGC 5273 \citep{bentz14, merrell22}, and MCG--06-30-15 \citep{bentz16a, das25}. 

In this work, we carry out new stellar-dynamical modeling of NGC 3227. 
\citet{davies06} were only able to constrain the mean velocity and velocity dispersion, $v$ and $\sigma$, from their observations; the S/N of the data did not allow them to constrain the higher-order Gauss-Hermite (GH) moments ($h_3 - h_6$) that more fully describe the line-of-sight velocity distributions (LOSVDs). These higher-order moments show the deviation of the LOSVDs from a pure Gaussian profile and are necessary to correctly model the orbital distribution of the stars. Without the higher order moments, stellar-dynamical modeling results are subject to the mass-anisotropy degeneracy \citep{binney82,gerhard93}.

Moreover, the inferred central $M_{BH}$ depends linearly on the distance to the source. \citet{davies06} used a redshift-based distance to the galaxy of $D= 17$ Mpc. Since NGC 3227 is at a very low redshift ($z$ = 0.00377), the dominant factor to its recessional velocity may not be the expansion of the universe, and therefore the redshift may not accurately reflect the distance.

In our work, we used archival $K$-band spectroscopy with higher S/N, and were able to constrain the higher order GH moments in the LOSVDs. We also adopt a distance of D=$23.7\pm2.6$ Mpc which was measured for the companion galaxy NGC 3226 using surface brightness fluctuations (\citealt{tonry01}, with the adjustments of \citealt{blakeslee01}). We note this distance is in agreement with the distance to NGC 3227 reported by \citet{robinson21} of $D=24.3\pm 4.9$ Mpc based on \ion{H}{1} 21\,cm observations and the \citet{tully77} relationship.

This paper is organized as follows. In Section \ref{sec:obs}, we discuss details of the observations. In Section \ref{sec:kin}, we present our kinematic measurements. Section \ref{sec:forstand} describes the parameters used for modeling the data, while our results are presented in Section \ref{sec:result}. We discuss the interpretations of our findings in Section \ref{sec:discuss} and finally present a summary in Section \ref{sec:sum}.

\section{Integral Field Spectroscopy} \label{sec:obs}
NGC 3227 is located at $\alpha = 155.8773^\circ$ and $\delta = 19.8650^\circ$.
Archival $K$-band spectroscopic observations from the Near-Infrared Integral Field Spectrograph (NIFS) on Gemini North were used for this work. The observations were taken on two nights in December 2016, with the intent of studying the feedback processes in this nearby AGN.
We obtained the data from the Gemini observatory archive under program no.\ GN-2016A-Q-6 (PI: Storchi-Bergmann). The observations were collected in a source--sky--source pattern ensuring that every source exposure has a sky exposure either immediately before or after. The K\_G5605 grating and HK\_G0603 filter were used in the observations. Further details about the observing program are described by \citet{riffel17}.

The NIFS reduction pipeline\footnote{https://github.com/jlwalsh12/NIFS-reduction-pipeline/} was used to reduce the data, following the recommendations of \citet{merrell20}.
Reduction steps included correction for bias and flats, dark subtractions, identifying bad pixels, wavelength calibration, correction for spectral and spatial distortions, and compensation for cosmic rays and negative pixels. We did not find any dark frames collected on the nights of observing, so we used the dark frames from the nights before. Telluric lines were corrected using a standard star (HIP 45874) spectrum. The pipeline also shifted the spectra to heliocentric velocities. 
A total of nine science observations from two nights with exposure times of 400 s each were co-added to create our final data cube, resulting in a total on-source time of 3600 s. The final cube is sampled at a resolution of $0\farcs05$/pixel and covers a field of view (FOV) of $2\farcs45 \times 2\farcs45$ after trimming.
We used Xe$-$Ar arc lamp calibration lines to measure the instrumental resolution to be 3.2 \AA.

\begin{deluxetable}{cccc}
\tabletypesize{\footnotesize}
\tablewidth{0pt}
 
 \tablecaption{PSF model \label{tab:psf}}
 \tablehead{
 \colhead{Component} & \colhead{Flux Contribution} & 
 \multicolumn{2}{c}{Width ($\sigma$)} \\
 \colhead{} & \colhead{(weight)} & \colhead{(arcsec)} & \colhead{(pc)}
  }
 
 \startdata 
        \hline
        1 & 0.536 & 0.073 & 8.3\\
        2 & 0.464 & 0.301 & 34.4\\
        \hline
 \enddata
 \tablecomments{The relative flux contribution (weight) and Gaussian widths of the components of the point spread function for the NIFS observations. At the distance to NGC 3227, $D=23.7$ Mpc, $1''$ corresponds to $114.3$ pc.}
\end{deluxetable}
We characterized the point spread function (PSF) using the broad $Br{\gamma}$ line of the AGN, which arises from the spatially unresolved broad line region and therefore acts as a point source. To create a PSF image, we took several slices from the wing of the broad line and an equal number of slices from the continuum near the broad line. We summed the broad-line image slices and then subtracted the sum of the continuum image slices. We then fit the resultant PSF image with a set of concentric circular Gaussian profiles using \texttt{Galfit} \citep{peng02,peng10}. The details of our PSF model are given in Table \ref{tab:psf}.

\begin{figure*}
    \centering
    \gridline{\fig{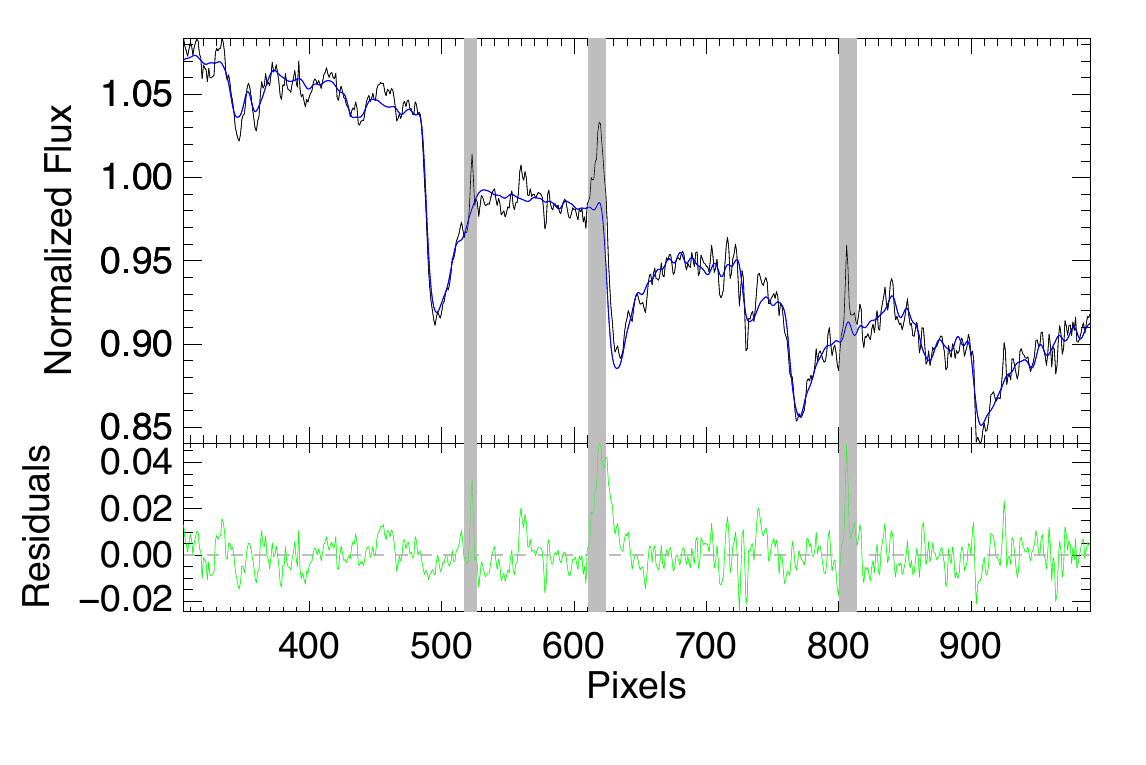}{0.492\textwidth}{}
           \fig{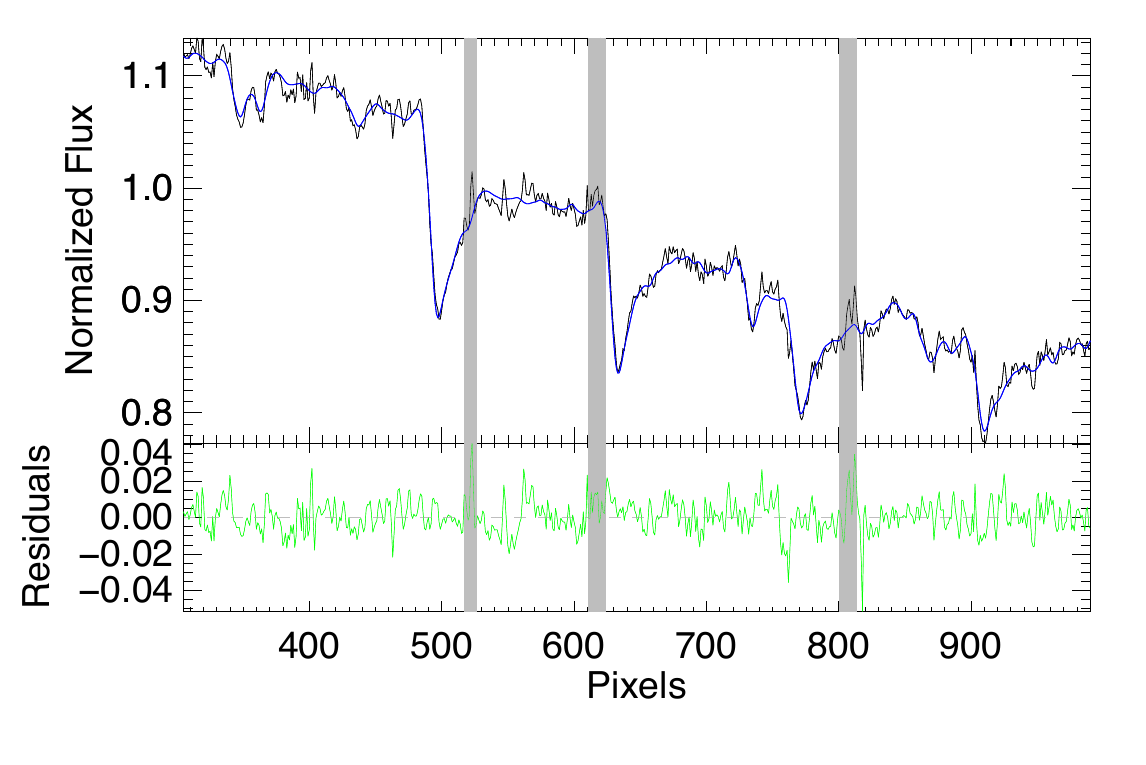}{0.492\textwidth}{}}
    \caption{Normalized flux vs.\ pixels for a typical single-spaxel bin (\textit{left}) and a multiple-spaxel bin (\textit{right}) in our data cube. The single-spaxel bin is $\sim 0\farcs05$ away from the center, while the multiple-spaxel bin is $\sim 0\farcs25$ away from the center. In the top panels, the observed data and the best-fit templates are shown in black solid lines and blue dashed lines, respectively. Bottom panels show the residuals as green solid lines. The grey vertical stripes denote masked-out pixels that were ignored for spectral fits.}
    \label{fig:spectrum}
\end{figure*}

\begin{figure*}
    \centering
    \includegraphics[width=\linewidth]{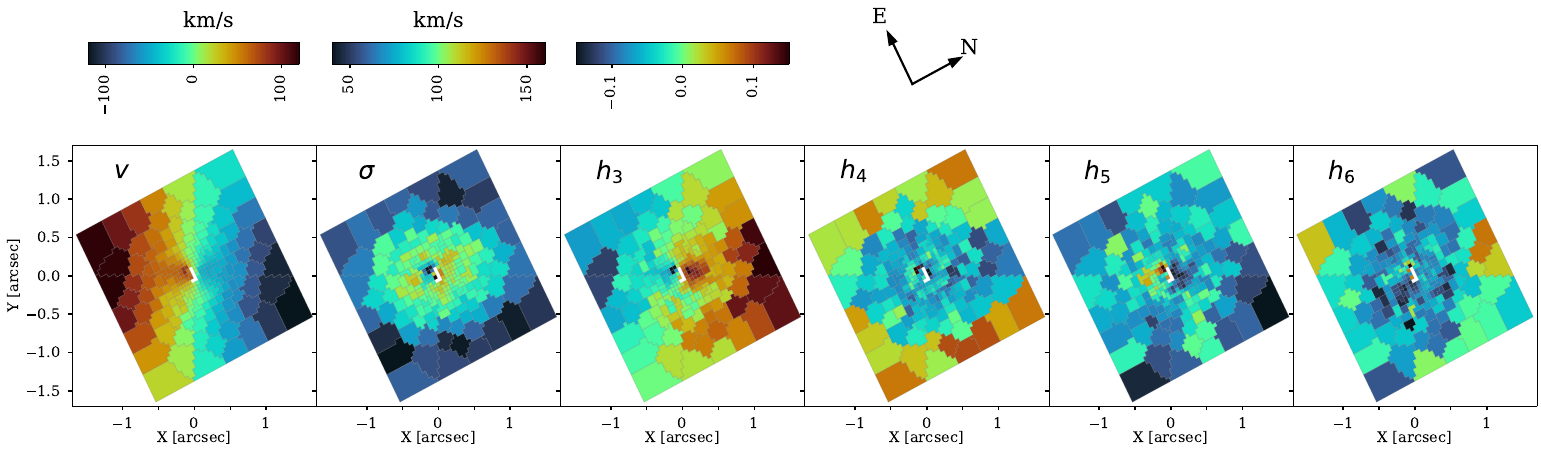}
    \caption{Kinematic maps of the LOSVDs based on Gemini/NIFS data. The panels are rotated so that the kinematic major axis is horizontal. The panels show velocity ($v$), velocity dispersion ($\sigma$), and higher-order GH moments ($h_n$; $3\le n\le6$). The ranges for $v$ and $\sigma$ are shown above the panels, while ($h_3 - h_6$) share the same scale. The white region at the center denotes the bins that were masked out during the dynamical modeling because of their poorly constrained kinematics due to the AGN dominating the continuum emission.}
    \label{fig:kinmap}
\end{figure*}

\section{Kinematics} \label{sec:kin}

We used \texttt{pPXF} \citep{cappellari04, cappellari17} to extract the bulk stellar kinematics in the nucleus of NGC 3227. Using the \texttt{fitkinpa} routine \citep{Kraj06}, we measured the kinematic position angle for NGC 3227 to be $150\pm 1$ degrees, East of North. We then binned the data cube using the kinematic major axis and the minor axis as axes of symmetry. This binning pattern enables point-symmetric fitting of the spectra. However, the nuclear stellar kinematics in NGC 3227 is not exactly axisymmetric, so we fitted the full plane.

For the binning pattern, we used Voronoi binning \citep{cappellari04}, where spaxels close to the galaxy center were treated as individual bins, and spaxels near the edge of the FOV where the signal is fainter are co-added into bins. This pattern ensured that high spatial resolution was maintained near the black hole while the S/N remained above some minimum threshold for every bin. After binning, we fit the spectrum in each bin and measured the LOSVDs. 

Stellar templates from the NIFS library \citep{Winge09} were used for fitting the binned spectra. The template library consists of 30 stellar spectra of  G, K, and M giant stars. We primarily aimed to fit the CO band-head signatures near 2.29 $\mu$m, so we selected a fitting window ranging from 2.20 -- 2.45 $\mu$m. Emission lines and noise spikes appearing in the fitting window were excluded using the \texttt{goodpixels} keyword in \texttt{pPXF}. To further improve our fitting of the continuum, we used additive and multiplicative Legendre polynomials in the \texttt{pPXF} routine. The additive polynomials take care of template mismatch, contribution from the AGN continuum \citep{cappellari17}, and poor sky subtraction, while multiplicative polynomials are needed to account for reddening and calibration errors. The orders of the Legendre polynomials were selected after comparing the spectral fits from different combinations of these polynomials, and choosing the lowest orders that minimize the residuals but do not overfit the continuum. In our final fit, we adopted 3rd order additive and 1st order multiplicative  polynomials. We further added a penalty value in \texttt{pPXF} that improves the LOSVD measurements by penalizing the noisy bins and drawing the higher-order ($h_3 - h_6$) moments closer to zero \citep{cappellari04}. To choose the penalty value, we compared the resultant maps when adopting a penalty value of 0.0-0.5 in steps of 0.1, choosing the lowest value that reduced the noise in the maps without oversmoothing the higher order moments. Our adopted penalty value was $0.1$ in our final fits.

After fitting all the spectra, we selected one of the bins with the lowest $\chi^2$, and adopted the template used to fit the spectrum in that particular bin as the `optimal template'. This optimal template was then used to fit all the binned spectra across the entire FOV. This step ensures that we maintain consistency of the stellar population within this small FOV and that we mitigate any possible biases that may arise from different metallicities of the stars, which may in turn affect the line shape measurements. Our `optimal template' is a weighted sum of eight stars of the following spectral types: G8III, K0III, K2III (two), K5Ib, M0.5Ib, M2III, and M3I.

Two examples of our final spectral fits are shown in Figure \ref{fig:spectrum}. The black  lines  denote the observed data and the blue dashed lines denote the best-fit template, with masked emission lines shown in grey. The residuals are plotted in green. By fitting the CO absorption profiles in the spectra, \texttt{pPXF} measures the LOSVDs of the nuclear stars in each bin using GH moments. By default, the first two moments, $h_1$ and $h_2$ are set to zero in \texttt{pPXF}. Mean velocities and velocity dispersions are instead calculated, along with the higher order moments ($h_3$--$h_6$) that describe the deviation of the LOSVD from a pure Gaussian profile. The odd moments $h_3$ and $h_5$ indicate asymmetric deviations similar to skewness in the LOSVDs while the even moments $h_4$ and $h_6$ capture the flatness or peakyness (similar to kurtosis) of the profiles. In Figure \ref{fig:kinmap}, we present the kinematic maps of the stellar LOSVDs. The maps are rotated to align the kinematic major axis with the horizontal axis. The central white regions in Figure \ref{fig:kinmap} denote a few rejected bins that were dominated by the AGN and were masked out during the modeling.

\begin{figure}
    \centering
    \includegraphics[width=\linewidth]{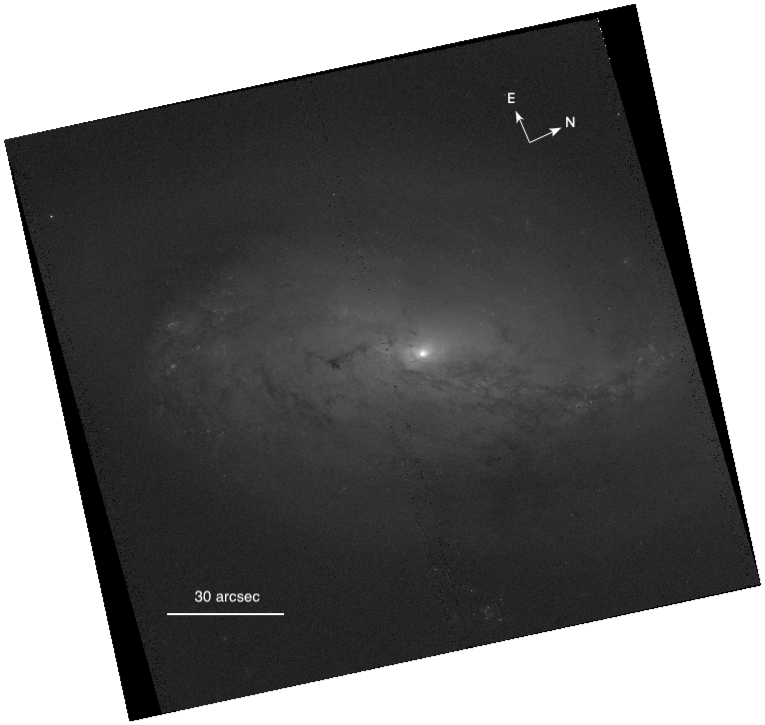}
    \caption{Logarithmically scaled HST WFC3 F547M image used for creating the surface brightness profile of NGC~3227, rotated to match the orientation of the kinematic maps in Figure \ref{fig:kinmap} (the kinematic major axis is aligned with the horizontal axis).}
    \label{fig:n3227}
\end{figure}

\begin{deluxetable}{ccccc}
 \tabletypesize{\scriptsize}
 \tablewidth{0pt}
 \tablecaption{Surface Brightness Profile\label{tab:surface}}
 \tablehead{
 \colhead{Components} & \colhead{Magnitude} & \colhead{Half-light radius} & \colhead{S$\text{\'e}$rsic index} & \colhead{Axis ratio}\\
 \colhead{   } & \colhead{$m_V$ (mag)} & \colhead{$r_e$ (arcsec)} & \colhead{$n$} & \colhead{$q^{obs}$}
  }
 
 \startdata 
        \hline
        nuclear disk & 16.38 &  0.35 & 1.21 & 0.45\\
        bulge        & 15.46 &  0.70 & 0.79 & 0.72\\
        bar          & 14.25 &  4.25 & 1.14 & 0.57\\
        disk         & 11.04 & 58.39 & [1.00] & 0.44\\
        \hline
 \enddata
 \tablecomments{Parameters of the best-fit surface brightness profile for NGC 3227. There are four components in this galaxy --- nuclear disk, bar, bulge, and a large scale disk. We characterized the components with $V$-band magnitude ($m_V$), half-light radius ($r_e$), S$\text{\'e}$rsic index ($n$), axis ratio ($q^{obs}$), and position angle (PA). Values held fixed are reported in brackets. The PA was the same for all four components and a best-fit value of $PA=-24.09$\degree\ East of North was found. The $V$-band magnitudes, $m_V$ are corrected for Galactic extinction of $A_V$=0.062 mag.}
\end{deluxetable}

\section{Dynamical Modeling} \label{sec:forstand}

We used \texttt{FORSTAND}, a flexible orbit superposition toolbox for analyzing dynamical models\footnote{\url{https://github.com/GalacticDynamics-Oxford/Agama/blob/master/py/example_forstand.py}}, to create the dynamical models of NGC 3227, implementing the \citet{schwarzschild79} orbit-superposition technique. 
First, a 3D luminosity profile is derived by deprojecting the surface brightness profile. This luminosity profile is then converted to a mass density using a stellar mass-to-light ratio for each model. To account for the uncertainties from the deprojection, several models are constructed by assigning different flattening/thickness values to the third dimension of the 3D density.
The resulting gravitational potential from the 3D density has contributions from the central supermassive black hole, which is treated as a point source, the stars, and a dark matter halo.
Within this potential, a large library of stellar orbits is integrated over many dynamical times. The orbit weights are then determined by comparing the modeled LOSVDs from the combination of orbits  (convolved with the PSF) to the observed LOSVDs and the surface brightness profile, solving a quadratic optimization problem with non-negativity constraints on orbit weights, fitting the kinematic data in the least-square sense. The $\chi^2$ statistic is recorded for every model, and confidence levels for \mbh\ are determined around the minima of $\chi^2$.
 
In the following, we describe the specific modeling parameters used in this work.

\subsection{Single value parameters}
\texttt{FORSTAND} includes a number of parameters that have a single value chosen by the user that remain constant for all models.  These include the distance to the galaxy (D=$23.7\pm2.6$ Mpc) as well as the kinematics, the PSF model, and the surface brightness profile. We discuss these parameters below.

\begin{itemize}
    \item \textbf{Kinematic constraints and number of orbits}: Our kinematic constraints come from the stellar kinematics described in Section \ref{sec:kin}. We used a GH parameterization to describe the nuclear stellar kinematics. The number of constraints is determined by the number of bins present in the FOV multiplied by the number of GH moments in the LOSVDs of each bin. In this work, we have 242 bins and 6 moments for each bin, resulting in a total of 1452 kinematic constraints. The number of stellar orbits ($N_{\rm orbits})$ can in principle be chosen arbitrarily, but \citet{valluri04} suggested that $N_{\rm orbits}$ should be $\ge 5\times$ the number of kinematic constraints because for a significantly low number of orbits the $\chi^2$ contours may be too narrow and consequently, the models may not accurately constrain the \mbh. We tested 20,000 orbits but ultimately increased the number of orbits to 50,000 to reduce the noise in the $\chi^2$ contours.

    \item \textbf{PSF model}: The LOSVD of each orbit is convolved with the instrumental PSF. We adopted the PSF model described in Table \ref{tab:psf} of Section \ref{sec:obs}. \texttt{FORSTAND} parameterizes the PSF as the Gaussian widths in arcseconds, and the fraction of the flux contributed by each of the Gaussian profiles.

    \item \textbf{Surface brightness profile}: Using \texttt{Galfit}, we modeled the surface brightness profile of NGC 3227 from an HST WFC3 F547M image, shown in Figure \ref{fig:n3227}. The AGN contribution was accounted for using a PSF model and we included a linear gradient in $x$ and $y$ to account for the background sky.
    We included four concentric photometric components for NGC 3227 --- a nuclear disk, a bulge, a bar, and a large scale galactic disk. All components are assumed to be axisymmetric in our analysis. The components are characterized by their $V$-band magnitude ($m_V$), half-light radius ($r_e$), S$\text{\'e}$rsic index ($n$), axis ratio ($q^{obs}$), and position angle (PA). We tied the PAs of all the components together to accommodate axisymmetric dynamical modeling. The sky PA for all the components is PA=  $-24.09$\degree\ East of North. Our best-fit surface brightness profile is described in Table \ref{tab:surface}. The magnitudes reported in Table \ref{tab:surface} are corrected for Galactic extinction of $A_V$=0.062 \citep{schlafly11}. 

    We note that while \texttt{Galfit} provides the best fits to the image when a bar is included, we are building axisymmetric models in this work.  Therefore in the following discussion of the dynamical modeling, we treat this component that we refer to as a ``bar"  as a flattened and elongated spheroidal component.

\end{itemize}

\subsection{Varying parameters} \label{sec:params_varying}
\texttt{FORSTAND} also includes a variety of parameters that are varied across the models. These include our parameters of interest---\mbh\ and mass-to-light ratio---as well as flattening values for the 2D photometric components, dark matter halo scale radius and circular velocity, and random seeds. We also tested the robustness of our models by creating several different realizations of the orbit integrations by varying the random seed parameter  in \texttt{FORSTAND} used for generating the initial conditions for the orbit library. In the following paragraphs, we further discuss the varying parameters used in our models.
 
\begin{itemize}
     \item \textbf{Stellar mass-to-light ratio ($\Upsilon$):} We vary stellar M/L in multiplicative steps of $2^{0.05}$ with a starting value of 3 \msun/\lsun\ which is a standard M/L for disk galaxies, where the typical range is 1--10 \msun/\lsun\ \citep{zhu23}.
    
    \item \textbf{\mbh\ values:} We constructed models with black hole masses between $0-2\times 10^7$ \msun\ in steps of $1 \times 10^6$ \msun. All the \mbh\ inputs are multiplied by $\Upsilon$ when creating the dynamical models. The boundaries are selected after carefully scanning through a larger range of \mbh.
    
    \item \textbf{Flattening of galaxy components -- nuclear disk ($q_n$), bulge ($q_{\rm bulge}$), bar ($q_{\rm bar}$), and large scale disk ($q_d$):}
    Our best-fit surface brightness profile has four photometric components: a nuclear disk, a bulge, a bar, and a large scale disk.
    
    We explored models with a range of flattening values for each of these components. We tested two different disk flattening ($q_d$) values: $0.1$ and $0.2$, which are common values for disk flattening in edge-on disk galaxies \citep{tully00, kautsch06}. For each of the two $q_d$ values, we also tested a nuclear disk flattening ($q_n$) of $0.1$ and $0.2$. 
    For the bulge, flattening values ($q_b$) were allowed to vary within a range of [0.1,1.0] in steps of 0.1. And finally, the bar flattening values ($q_r$) were explored within [0.1,0.5] in steps of 0.1.

    We note that the adopted flattening of the disk sets the inclination, and in principle the other components' axis ratios as well, but we took the liberty to override these values manually, while rescaling the major axis lengths such that the circularly-averaged light profile of each component remains independent of its adopted flattening. The motivation is to account for uncertainties in the photometric modeling and propagate them to the dynamical models.
    
    \item \textbf{Dark matter halo}: In previous studies \citep{roberts21, merrell22, das25}, we have found that the small field of view of NIFS is not sufficient to constrain the dark matter halo without including larger-scale kinematic data, however, we  explored the influence of dark matter on our models here for completeness.  \texttt{FORSTAND} includes an NFW-type dark matter halo \citep{navarro96} that is defined by the peak circular velocity ($v_h$) and the scale radius ($r_h$). We calculated the $r_h$ value from the halo mass-concentration relation of dark matter \citep{dutton14}, and the stellar mass -- halo mass relation of \citet{behroozi19}. Based on these relations, we used a dark matter halo scale radius, $r_h =$ 13.7 kpc. Keeping the scale radius a constant, we explored circular velocities of $v_{h} = [100\sqrt{\Upsilon}, 200\sqrt{\Upsilon}, 300\sqrt{\Upsilon}] $ km/s. These values of $v_h$ were adopted to generously cover the typical values of circualr velocities in disk galaxies \citep{deisidio24}.
     
    \item \textbf{Random seed:} The random seed is used to set different initial conditions for the orbit library. We explored five different orbit realizations by varying the random seed parameter for our models.

\end{itemize}

\section{Results} \label{sec:result}

\begin{figure*}
    \centering
    \includegraphics[width=\textwidth]{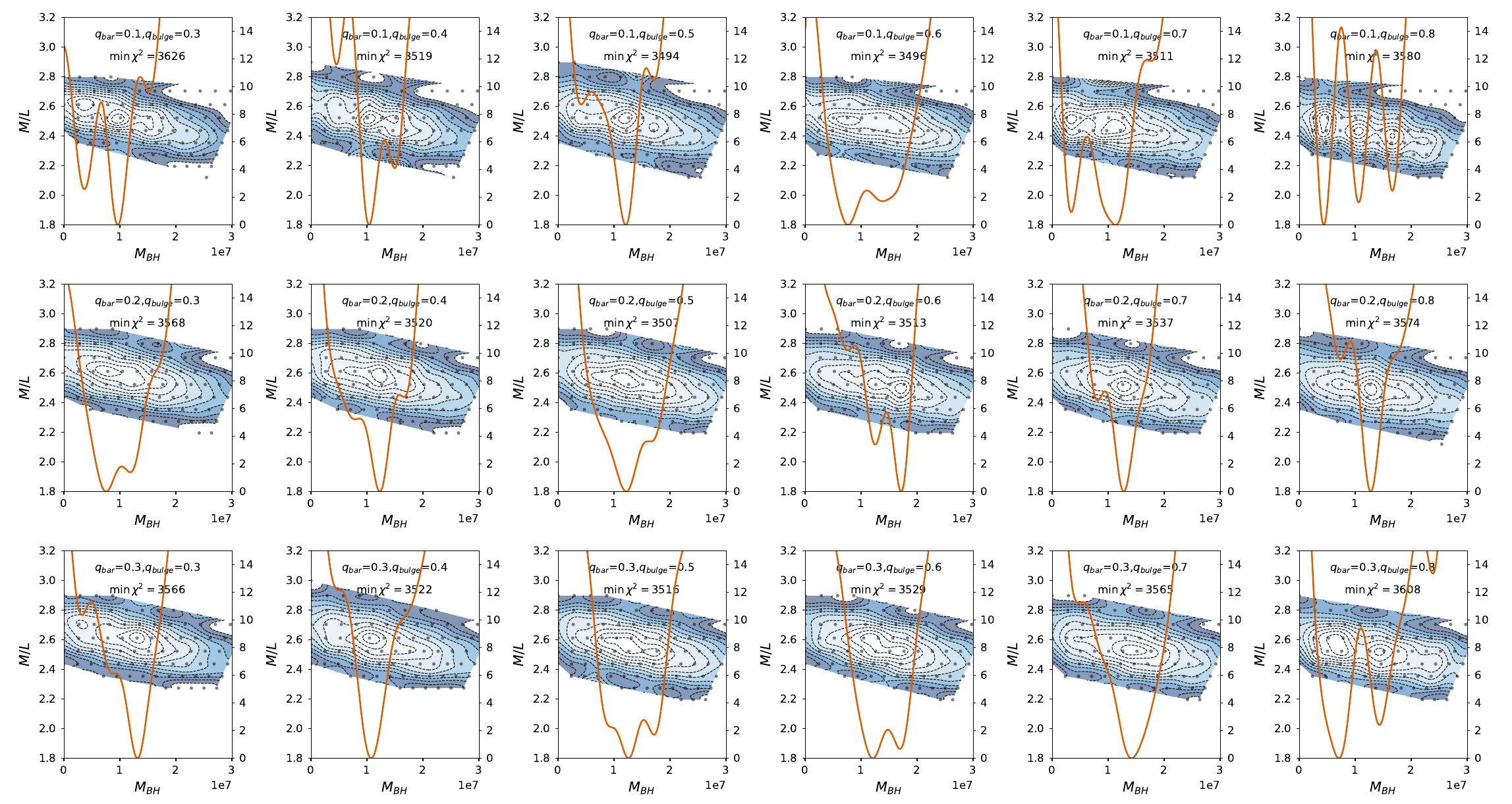}
    \caption{Example $\Delta\chi^2 = \chi^2 - \chi^2_{min}$ grid in the \mbh--M/L plane for different flattening values of the galaxy components. The secondary Y-axis shows the limit of the $\Delta\chi^2$. Each grey dot indicates a model. The dashed contour lines along with the blue 2D density maps indicate the $1\sigma, 2\sigma, 3\sigma,$ … confidence levels of 2D $\Delta\chi^2$ at 2.3, 6.2, 11.8, …, etc. for \mbh\ and M/L. The orange lines show the 1D $\Delta\chi^2$ value as a function of \mbh\ marginalized over M/L. The intrinsic flattening of the bar ($q_{\rm bar}$)  is fixed across the rows, whereas the intrinsic flattening of the bulge ($q_{\rm bulge}$) is fixed down the columns. The intrinsic flattening of the nuclear disk ($q_n$) and the large scale disk ($q_d$) are held fixed at 0.1 and 0.2, respectively, for all of the panels. In each panel, $q_{\rm bar}$, $q_{\rm bulge}$, and the minimum $\chi^2$ values are printed.}
    \label{fig:chi2}
\end{figure*}

We constructed many dynamical models for NGC 3227 by systematically varying the flattening of the four galaxy components -- nuclear disk, bulge, bar, and larger scale disk. An example of our results is shown in Figure \ref{fig:chi2}. Each panel in Figure \ref{fig:chi2} shows a plane of \mbh\ and stellar M/L for a specific combination of galaxy component flattening values. From the discrete grid of models, we have interpolated the $\chi^2$ values in the \mbh--M/L plane using a Gaussian Process (GP) regression to create the smooth 2D confidence levels and marginalized $\Delta\chi^2$ surfaces. The grey dots denote each computed model, the blue contours denote the 2D confidence levels for \mbh\ and M/L, and the orange curves denote 1D confidence levels for \mbh\ after marginalizing over all possible M/L values. The overall trend is that \mbh\ and M/L are anti-correlated. For the panels shown in this Figure, the nuclear disk flattening ($q_n$) and the outer disk flattening ($q_d$) are held constant at 0.1 and 0.2, respectively, and the different panels show varying flattening of the bulge ($q_{\rm bulge}$) and the bar ($q_{\rm bar}$). The bar flattening varies across the rows within the range [0.1--0.3] in steps of 0.1, and the bulge flattening varies across the columns within the range [0.3--0.8], also in steps of 0.1. 
 
We created five random realizations of our orbit library to test the robustness of the results, and averaged our results over all random seed values. The difference in $\chi^2$ values between these realizations is at the level $\mathcal O(10)$, which we take as a threshold of statistical significance. We also constructed models including a contribution from the dark matter halo but found that those models produced higher minimum $\chi^2$ values and worse fits overall.  

The best-fit models are chosen by goodness of fit from all of our dynamical models. Small differences on the order of a few $\times 10$ in the minimum $\chi^2$ are not statistically significant so we do not select a single panel as the best-fit model, but rather a family of best-fit models that includes several combinations of parameters that produce similar minimum values of $\chi^2$. Most of the best-fit models are found for $q_n=0.2$ and $q_d=0.1$ with $q_{\rm bar}=[0.1-0.2]$, $q_{\rm bulge} = [0.4-0.7]$, and $q_{\rm bar}=0.3$, $q_{\rm bulge}=[0.4-0.5]$ (see Figure \ref{fig:chi2}). A few additional best-fit models were found for $q_n=0.1$ and $q_d=0.1$ with $q_{\rm bar}=0.2$, $q_{\rm bulge}=[0.5-0.6]$.

\begin{figure*}
    \centering
    \includegraphics[width=\textwidth]{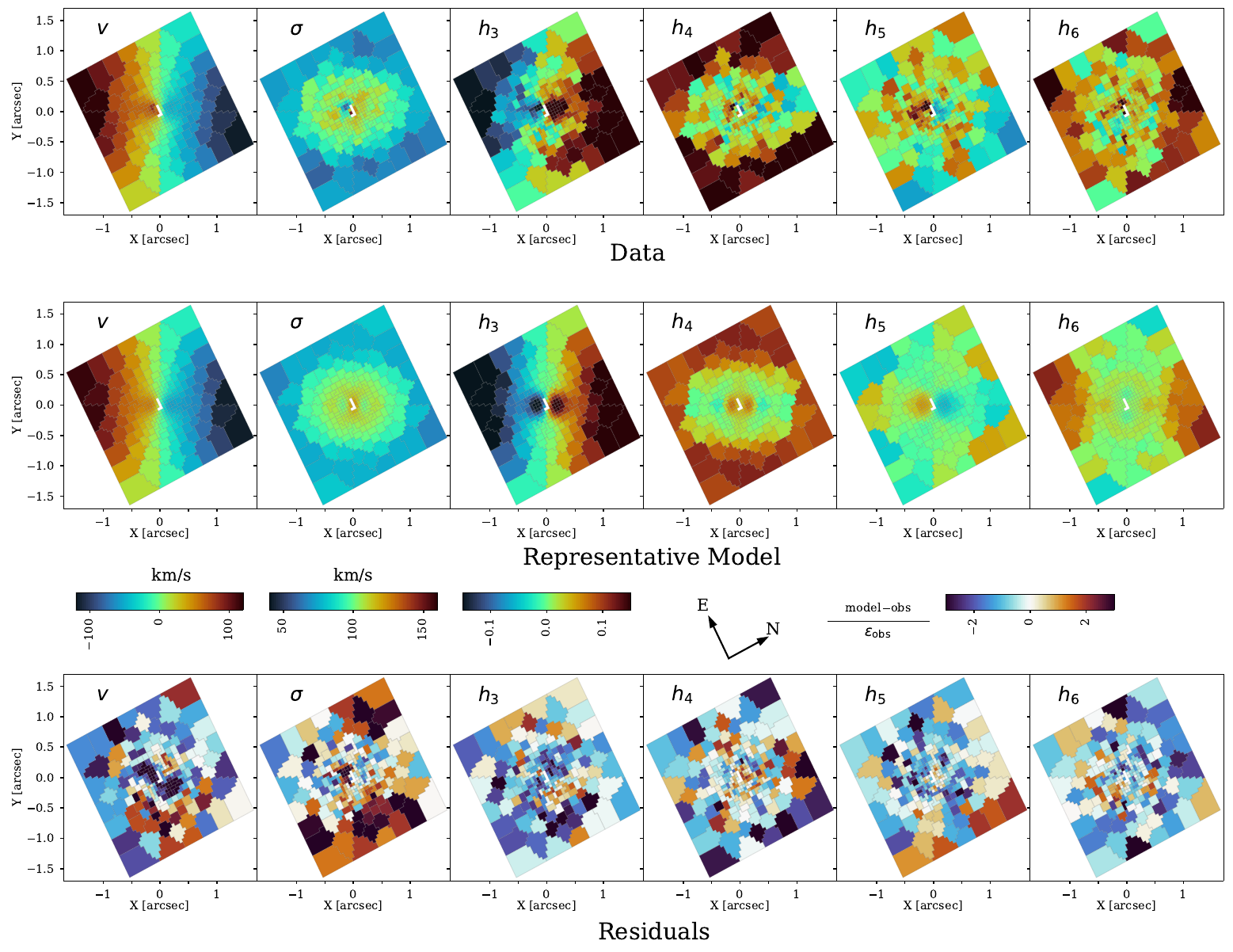}
    \caption{Comparison of kinematic maps between the data and a representative model, along with the residuals. The white region at the center denote the bins that were ignored for dynamical modeling because of their poorly constrained kinematics due to the AGN dominating the continuum emission. The color maps are the same as Figure \ref{fig:kinmap}.}
    \label{fig:kinmap_compare}
\end{figure*}

From our best-fit models, we selected a representative model to compare the observed kinematics with. The parameters for the representative model are $q_n=0.2$, $q_{\rm bulge}=0.5$, $q_{\rm bar}=0.2$, and $q_d=0.1$. A comparison between the observed kinematics and the kinematics from the representative model is presented in Figure \ref{fig:kinmap_compare}. The top panel shows the full-plane kinematic maps from the data (same as Figure \ref{fig:kinmap}). The kinematic maps from the representative model are shown in the middle row, and the normalized residuals are plotted in the bottom row. The residuals are calculated by subtracting the observed kinematics from the model kinematics, normalized by the observed uncertainty ($\epsilon_{obs}$) in the LOSVDs. All maps are rotated to show the maximum rotation (kinematic major axis) along the horizontal axis. The representative model prefers slightly higher values of $\sigma$ and lower values of $h_4$ along the edges of the FOV compared to the observations. Therefore, we see noticeable residuals in the $\sigma$ and $h_4$ maps.

\begin{figure*}
    \centering
    \includegraphics[width=0.49\textwidth]{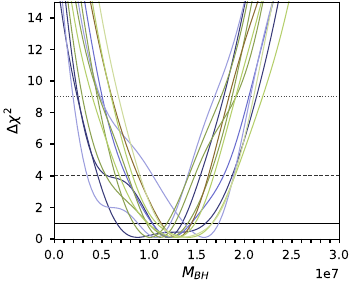}
    \includegraphics[width=0.49\textwidth]{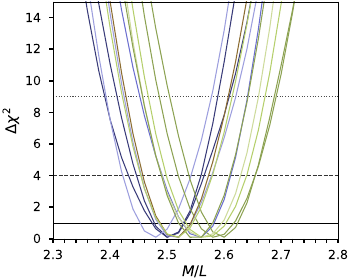}
    \caption{Marginalized $\Delta\chi^2$ plots for \mbh\ and M/L. The different lines correspond to all of our best-fit models. The darker the shade of the line, the lower their minimum $\chi^2$ value is. 1$\sigma$, 2$\sigma$, and 3$\sigma$ confidence levels are marked by solid, dashed, and dotted horizontal lines respectively.}
    \label{fig:final}
\end{figure*}

We note that the minimum $\chi^2$ values from all best-fit models in Figure \ref{fig:final} are significantly higher than the number of kinematic constraints; this may suggest that the observed uncertainties are somewhat underestimated, but there are also clear large-scale patterns in the model residuals, such as the discrepancies in $\sigma$ and $h_4$ described above for the representative model. Since $\chi^2$ per data point is well above unity, a standard statistical criterion of $\Delta\chi^2=1$ defining the 1-$\sigma$ uncertainty interval on model parameters would be too restrictive. Moreover, as discussed above, the values of $\chi^2$ for the same model have some intrinsic scatter at a level $\mathcal O(10)$ due to the discreteness of the orbit library, so the location of the true minimum of $\chi^2$ is hard to determine precisely. Instead, we take the spread in the locations of the $\chi^2$ minima among all models declared to be sufficiently good fits as a more reliable uncertainty measure than the formal 1$\sigma$ intervals corresponding to $\Delta\chi^2=1$ for one best-fit family of models (see, e.g., Section~6.3 in \citealt{merrell22}). As an additional sanity check, we ran models using only the kinematic data from the left or the right half of the IFU (while the models are symmetric by construction, they were fitted to the input data that were intentionally not symmetrized). We found that the results from modeling the full-plane kinematics include the range of results that were found using each half-plane separately.

The best-fit models give similar, though slightly different, constraints on \mbh\ and stellar M/L. To determine the final, combined constraints on \mbh\ and M/L, we overplotted the marginalized $\Delta\chi^2$ curves for these two parameters from all of our best-fit models, as presented in Figure \ref{fig:final}. The left and right panels display the marginalized 1D $\Delta\chi^2$ for \mbh\ and M/L, respectively. The colors show the difference between the minimum $\chi^2$ value from that particular model and the lowest $\chi^2$ across all models. Darker shades indicate smaller difference and lighter shade indicate larger difference. The overlapping region from all the best-fit models determines the error bars for \mbh\ and M/L. The solid, dashed, and dotted lines denote the 1$\sigma$, 2$\sigma$, and 3$\sigma$ confidence levels for these two parameters. From our best-fit models, we measure a black hole mass of $(1.2\pm0.5)\times 10^7$ \msun\ for NGC 3227, and a $V$-band stellar mass-to-light ratio of $2.53\pm0.09$ \msun/\lsun. 

\section{Discussion} \label{sec:discuss}

As previously mentioned, \citet{davies06} measured a stellar-dynamical mass for this black hole using data from SINFONI on VLT, implementing the Schwarzschild orbit superposition method. The S/N in their data was too low to constrain the higher-order moments in the LOSVDs beyond velocity and velocity dispersion. These higher-order moments, however, help break the mass-anisotropy degeneracy \citep{gerhard93}. While \citet{davies06} allowed for a large range of possible values for $h_3$ and $h_4$ to avoid this bias, their derived black hole mass of $M_{BH} = (0.7-2.0) \times 10^7$ \msun\ may carry additional uncertainties beyond the quoted error bars given the limitations of the data.  Additionally, they used a redshift-based distance to NGC 3227. Since the redshift of NGC 3227 at $z = 0.00377$ is very low, its recessional velocity is not necessarily dominated by the Hubble flow. In this work, we adopted a  distance of $D=23.7\pm2.6$ Mpc, measured for the interacting companion galaxy NGC~3226. Since distance and dynamical mass are linearly correlated, we scaled the \mbh\ from \citet{davies06} before comparing with our results, finding a revised stellar-dynamical mass of $M_{BH} = (1.9\pm0.9) \times 10^7$ \msun. Using higher S/N Gemini NIFS data we constrained the higher-order moments of the stellar LOSVDs and measured \mbh=$(1.2\pm0.5) \times 10^7$ \msun\ for the SMBH in NGC 3227, which is consistent with but more tightly constrained than the \mbh\ derived by \citet{davies06}.

NGC 3227 is one of the rare AGN that have multiple direct \mbh\ constraints. \citet{hicks08} constructed gas-dynamical models by studying the circumnuclear disk in NGC 3227 with AO-assisted NIRSPEC observations from Keck, and measured  $M_{BH} = (2.0^{+1.0}_{-0.4}) \times 10^7 M_{\odot}$.
However, they also used a different redshift-based distance to this galaxy. The distance-corrected gas-dynamical mass  is $M_{BH} = (3.1^{+1.5}_{-0.6}) \times 10^7$ \msun, which is somewhat larger than the value we report here, though they are within 2$\sigma$ of each other. 

NGC 3227 has also been a successful RM target  \citep{denney10, derosa18, brotherton20,bentz23b}. RM masses for the black hole determined from the average time delay and typical width of the broad H$\beta$ emission line range from $2.3\times10^6 - 1.6\times10^7$ \msun\  when a scale factor of $\langle f \rangle=4.82$ \citep{batiste17b} is adopted. However, the scale factor is a population average and is expected to vary from AGN to AGN (e.g., \citealt{villafana22}), and may also vary on a dynamical timescale within a single AGN. More recently, dynamical modeling of the velocity-resolved reverberation response in NGC 3227 provided a new mass constraint independent of the adoption of any scale factor, \mbh = $(1.23^{+1.52}_{-0.67}) \times 10^7$~\msun\ \citep{bentz23b}, which is in good agreement with our best-fit \mbh\  from stellar-dynamical modeling.

Black hole masses for NGC 3227 from all  three techniques are presented in Table \ref{tab:mbh} for comparison. The stellar- and gas-dynamical masses are scaled to assume the same distance, $D=23.7\pm2.6$ Mpc. It should be noted that the \mbh\ measurement from reverberation mapping is distance--independent. In Table \ref{tab:mbh}, we only report the RM mass based on modeling the velocity-resolved reverberation response since it does not rely on the adoption of a scale factor that may introduce additional uncertainties.

\citet{davies06} modeled two separate contributions to the total stellar mass---a nucleus and a bulge---each with their own independent $K$-band M/L. From their models, they found M/L$=1-5$ \msun/\lsun\ for the nucleus and $25-35$ \msun/\lsun\ for the bulge. In our modeling, we constrained a single M/L for the combination of the four surface brightness components determined from high-resolution optical HST imaging.  The differences in modeling approach as well as photometric band makes it difficult to directly compare our best-fit M/L with those reported by \citet{davies06}.  However, our best-fit $V$-band M/L of $2.53\pm0.09$ \msun/\lsun\  lies within the range of typical mass-to-light ratios found for other galaxies with stellar masses of $\sim 10^{11}$ \msun\ \citep{zhu23}, as has been found for NGC 3227 \citep{bentz18}.

\begin{deluxetable}{ccc}
\tabletypesize{\footnotesize}
\tablewidth{0pt}
 
 \tablecaption{\mbh\ comparison \label{tab:mbh}}
 \tablehead{
 \colhead{Technique} & \colhead{\mbh\ ($10^7$ \msun)} & \colhead{Reference}
  }
 
 \startdata 
        \hline
        stellar dynamics & $1.2\pm 0.5 $ & this work\\[1pt]
        stellar dynamics & $1.9\pm0.9$ & 1\\[1pt]
        gas dynamics & $3.1^{+1.5}_{-0.6}$  & 2 \\[1pt]
        velocity-resolved RM &  $1.23^{+1.52}_{-0.67}$ & 3\\[1pt]
        \hline
 \enddata
 \tablecomments{All black hole masses for NGC 3227 from  direct \mbh\ measurement techniques. The stellar- and gas-dynamical masses are scaled assuming $D=23.7$ Mpc. References: 1. \citet{davies06}, 2. \citet{hicks08}, 3. \citet{bentz23b}}
\end{deluxetable}

\section{Summary} \label{sec:sum}

In this work, we present a revised stellar-dynamical mass for the central supermassive black hole in NGC 3227. 
We used archival NIFS data from Gemini to measure the LOSVDs of the nuclear stars in this galaxy. We constructed dynamical models using the open-source code \texttt{FORSTAND}, implementing the Schwarzschild orbit-superposition technique. Our best-fit models indicate a central black hole mass of \mbh=$(1.2\pm0.5) \times 10^7$ \msun\ in NGC 3227.
Our results are consistent with the stellar-dynamical \mbh\ from \citet{davies06}, but are more tightly constrained. They are also consistent with the mass derived from modeling the velocity-resolved reverberation response in NGC 3227 by \citet{bentz23b}. On the other hand, our stellar-dynamical mass is  slightly smaller than the gas-dynamical mass reported by \citet{hicks08} (after scaling to the same distance), but they are within 2$\sigma$ of each other.
We also constrained the central $V-$band stellar mass-to-light ratio to be $M/L = 2.53\pm0.09$ \msun/\lsun\ in NGC 3227, which is a typical value for a galaxy with a stellar mass of $\sim 10^{11}$ \msun.

Our new stellar-dynamical mass provides a valuable consistency check for the mass of the SMBH in NGC 3227. Such comparisons are necessary to put all black hole masses on the same scale, and to enrich our understanding of  black hole-galaxy co-evolution across cosmic time.
\begin{acknowledgments}
N.\ Das and M.C.\ Bentz gratefully acknowledge support from the NSF through grant AST-2407802. EV acknowledges support from an STFC Ernest Rutherford fellowship (ST/X004066/1).
The work presented here is based on observations from the Gemini Data Archive.
IRAF was distributed by the National Optical Astronomy Observatory, which was managed by the Association of Universities for Research in Astronomy (AURA) under a cooperative agreement with the National Science Foundation.
\end{acknowledgments}

\begin{contribution}

N.\ Das reduced the NIFS data, extracted the kinematics, created the dynamical models, and led the writing of the manuscript.
M.C.\ Bentz oversaw the project and carried out the surface-brightness profile fitting.
E.\ Vasiliev and M.\ Valluri provided the dynamical modeling code, as well as expertise in constructing the models and interpreting the findings.
C.A.\ Onken provided subject matter expertise.
All authors contributed to the presentation of the results.

\end{contribution}

\facilities{Gemini (NIFS), HST(WFC3)}

\software{pPXF \citep{cappellari04, cappellari17}, Galfit \citep{peng02,peng10}, FORSTAND \citep{vasiliev20}
          }

\bibliography{ND_ref}{}
\bibliographystyle{aasjournalv7}

\end{document}